# A methodology to compute the territorial productivity of scientists: The case of Italy[1]

Giovanni Abramo (corresponding author)
*Laboratory for Studies of Research and Technology Transfer*
*Institute for System Analysis and Computer Science (IASI-CNR)*
*National Research Council of Italy*
    ADDRESS:    Istituto di Analisi dei Sistemi e Informatica
                          Consiglio Nazionale delle Ricerche
                          Via dei Taurini 19, 00185 Roma – ITALY
                          tel. +39 06 7716417, fax +39 06 7716461
                          giovanni.abramo@uniroma2.it

Ciriaco Andrea D'Angelo
*University of Rome "Tor Vergata" - Italy and*
*Laboratory for Studies of Research and Technology Transfer (IASI-CNR)*
    ADDRESS:    Dipartimento di Ingegneria dell'Impresa
                          Università degli Studi di Roma "Tor Vergata"
                          Via del Politecnico 1, 00133 Roma – ITALY
                          tel. and fax +39 06 72597362
                          dangelo@dii.uniroma2.it

**Abstract**

Policy-makers working at the national and regional levels could find the territorial mapping of research productivity by field to be useful in informing both research and industrial policy. Research-based private companies could also use such mapping for efficient selection in localizing R&D activities and university research collaborations. In this work we apply a bibliometric methodology for ranking by research productivity: i) the fields of research for each territory (region and province); and ii) the territories for each scientific field. The analysis is based on the 2008-2012 scientific output indexed in the Web of Science, by all professors on staff at Italian universities. The population is over 36,000 professors, active in 192 fields and 9 disciplines.

**Keywords**

*Spatial analysis; regional innovation systems; research productivity; universities; bibliometrics*

---

# 1. Introduction

The localization of universities within a particular nation has historic, economic, and sociological origins, and more recently is ever more influenced by policy and strategic decisions. Whatever the origin of the current territorial distribution of new knowledge suppliers, the policy maker certainly has interests in monitoring the evolution of the efficiency of research activities, for purposes of understanding and decision-making regarding the selective allocation of public resources. Similarly, for research-based companies in the private sector, the territorial mapping of research productivity by field can inform efficient choices in the localization of R&D activities, and research collaborations.

In the literature, the characterization of the scientific profile of a given territory is typically conducted by gathering and analyzing bibliometric data: specifically by analyzing the geographic distribution of scientific production, as indexed in the major bibliometric databases. This approach assumes that scientific publication in international journals is the principal form of dissemination of results from research activity, as conducted by universities and research institutions in general. Frenken et al. (2009) offer a particularly useful review of the full range of scientometric studies analyzing the spatial dimension of scientific production, beginning from the pioneering works by Narin and Carpenter (1975) and Frame et al. (1977). This latter work, under the suggestive title "The distribution of world science", and based on data from the ISI Science Citation Index, maps the distribution of output from 117 countries and in 92 disciplines, over one year (1973). More recent studies, employing similar methodologies, have primarily concerned the spatial concentration of scientific production, which seems to have remained high for the industrialized nations of the OECD. These nations thus continue to account for the major share of world output (May, 1997; Adams, 1998; Cole and Phelan, 1999; Glänzel et al., 2002; King, 2004; Horta and Veloso, 2007), despite a rapid increase in scientific production from China (Leydesdorff and Zhou, 2005). Analyses at the regional level have been less frequent: one case is the work by Matthiessen and Winkel-Schwarz (1999), on the analysis of aggregated publication records for European metropolitan areas, for the years 1994-1996. Some scholars have also proposed analyses based on the spatial distribution of highly-cited publications, primarily for the identification of centers of excellence at the regional level (Bonitz et al., 1997; Batty, 2003). More recently, a work by Bornmann and Leydesdorff (2011), based on the Web of Science (WoS) data, identifies cities where top-10% highly-cited papers were published more frequently than would be expected, offering visualization of the results via Google Maps. In very similar manner, Bornmann et al. (2011) present methods for mapping centers of excellence around the world, in this case using Scopus data. Excellence in single scientific fields is identified, revealing agglomerations in regions and cities where highly-cited papers (top-1%) were published. Shifting the focus from cities to regions, Bornmann and Waltman (2011) use visualization methods (density maps) to detect regions of excellence at the global level, focusing on the top 1% of 2007 papers indexed in Scopus. Very recently, Bornmann et al. (2014) presented a web application to identify research centers of excellence by field worldwide, using publication and citation data.

Within Italy, Tuzi (2005) pioneered bibliometric measures of the scientific specialization of regions, by two separate indicators: one based on publications and the other on average citations per paper. Morettini et al. (2013) document "knowledge activities" at the regional level, through the measurement of R&D expenditures, patents, and publications originating from "local labor systems". Abramo et al. (2009) have mapped the centers of excellence in Italy by analyzing the concentration of top scientists in the same institution. The same authors have recently presented a bibliometric methodology to carry out a spatial analysis of the impact produced by research institutions (Abramo et al., 2015), and to identify the scientific specialization of territories (Abramo et al., 2014b). The first contribution ranks territories in each research field by the total impact produced by local institutions, while the second one measures the "scientific" comparative advantages of territories. Both are held to have broader significance, most notably for the methodological approach. The authors use field-normalized citations, and not simply the counting of publications, to map the territorial distribution of new knowledge produced and the scientific specialization of regions: in fact, counts of publications alone do not permit an assessment of the real value of the new knowledge produced.

Continuing from their preceding works, these same authors now intend to apply bibliometric tools to measure the "scientific" competitive advantages of territories. We do so by measuring the territorial research productivity of professors. While the previous studies mapped the total amount of research impact produced at territorial level, this study maps the research productivity, i.e. the average impact per researcher. Research productivity, which is an efficiency measure of production, tells the potential research funders where research spending has the highest (or lowest) returns. The analysis simultaneously reveals: i) for each territory, the fields of research where the productivity is highest or lowest and ii) for each scientific field, which are the territories with the highest or lowest productivity. Findings of this type can inform public research and industrial policies at the national and regional levels. The so-called endogenous approach to local and regional development policy, is in fact based on the idea that regional development and the resulting economic growth is driven by endogenous forces in the form of "a highly educated workforce and knowledge and technologies developed in the region" (Todtling, 2010). A regional innovation system is conceived as involving various organizations concentrated in a geographical area - such as universities, public research institutions, companies and agencies active in technology transfer - which create, disseminate and apply new knowledge through interactive, cooperative activity. The region thus becomes the promoter of its own development and is seen as the territorial level of reference to engage growth, through local knowledge spillovers, intra-regional networks and labor mobility (Martin and Sunley, 1998; Krugman, 1991). At the regional level, universities are considered as the core knowledge-producing bodies, capable of a primary role in activating the innovation and development agenda, through their placement and the production of new knowledge for the industrial sector (Kitagawa, 2004; Thanki, 1999; Garlick, 1998; Foray and Lundvall, 1996). In addition to informing research policies, the territorial mapping of research productivity can also inform R&D labs localization strategies of hi-tech companies, and the choice of research collaboration with academia. It can also prompt further analyses aimed at delving into the causes of productivity differences, which can be due to historical reasons or structural reasons, such as the varying concentration of private R&D in specific fields.

To exemplify the application of the tool, we apply it to the analysis of the scientific competitive advantages of Italian territories, for which we have access to bibliometric data, disambiguated at professor's level, and to a database of the university affiliation of each professor. Beginning from the publications indexed in WoS between 2008 and 2012 and produced by professors on staff at Italian universities, the indicator "research productivity" is calculated. Its measure is based on the fractional counting of standardized citations received by indexed publications, in a manner taking due account of both the quantity and the impact of the scientific production from the researchers situated in a given territory.

The next section of the study presents the methodology, the dataset and the bibliometric indicator used for measuring productivity. The third and fourth sections present the results of the application of the methodology to the Italian territories. Section five concludes the work with the authors' comments.

2. Methodology

The steps of the methodology adopted in this work are the following: i) sorting of Italian universities by territory; ii) identification of all professors of each university and their ranking per research field; iii) identification of publications of each professor in the period 2008-2012; iv) measure of the research productivity of each professor; v) measure of the average research productivity of all professors belonging to the same territory and research field.

The Italian Ministry of Education, Universities and Research (MIUR) recognizes a total of 96 universities as having the authority to issue legally-recognized degrees. Twenty-nine of these are private, small-sized, special-focus universities. Sixty-seven are public and generally multi-disciplinary universities, scattered throughout the nation. In the overall system, 94.9% of faculty are employed in public universities. In keeping with the "Humboldt" university model, all professors are contractually obligated to carry out research, thus there are no teaching-only institutions in Italy. National regulations establish that each faculty member must allocate a minimum of 350 hours per year to teaching activities, of which no less than 120 to teaching classes. Public universities are largely financed by government, essentially through non-competitive allocation. Until 2009 the core government funding (56% of universities' total income) was input oriented, i.e. independent of merit, and distributed to universities in a manner intended to equally satisfy the needs of each and all, in function of their size and disciplines of research. It was only starting from 2009, following the first national research evaluation exercise (VTR), conducted between 2004 and 2006, that a minimal share, equivalent to 3.9% of total income, was assigned by the MIUR in function of the assessment of research and teaching. In the period of observation of this work, we can assume that no university was notably favored in terms of public funds allocation. Because there is no reason to expect that the very few variabilities in teaching load or public funds allocation are concentrated in few particular regions, differences in research productivity can then be the result of differences mostly in merit and/or in private sector funding. The latter represents though a relatively small share of universities total income (ANVUR, 2014), and depends itself on merit but also on geographical proximity.

**2.1 Dataset**

Data on Italian academics in the observed period are extracted from the official database maintained by the MIUR[2]. The database indexes names, academic rank, affiliation, and research field of all academics in Italian universities (around 60,000 professors). In fact in Italy the MIUR manages a system for the classification of all professors into a total of 370 "scientific disciplinary sectors" (SDSs)[3]. Each professor belongs to one and only one of the SDSs, which are grouped into 14 university disciplinary areas (UDAs). Nine of the UDAs fall in the so-called hard sciences[4]. For reasons of robustness of bibliometric measures, we examine only those SDSs of the hard sciences in which at least 50% of the professors achieved at least one publication during the observed period. They are 192 SDSs out of 205.

The scientific output of each professor is extracted from the Italian Observatory of Public Research (ORP), a database developed and maintained by the authors and derived under license from the Thomson Reuters WoS. Then by applying a complex algorithm for disambiguation of the true identity of the authors and their institutional affiliations (D'Angelo et al., 2011), each publication is attributed to the university professors that authored it, with a harmonic average of precision and recall (F-measure) equal to 96 (error of 4%). We further reduce this error by manual disambiguation. The dataset of the analysis is shown in Table 1.

*Table 1: Dataset for the analysis: number of fields (SDSs), universities, research staff and WoS publications (2008-2012) in each UDA under investigation*

| UDA | SDS | Universities | Research staff | Publications* |
|---|---|---|---|---|
| Mathematics and computer science | 10 | 69 | 3,387 | 16,920 |
| Physics | 8 | 64 | 2,497 | 23,587 |
| Chemistry | 12 | 61 | 3,174 | 26,703 |
| Earth sciences | 12 | 47 | 1,199 | 6,148 |
| Biology | 19 | 66 | 5,198 | 34,399 |
| Medicine | 50 | 64 | 10,966 | 71,575 |
| Agricultural and veterinary sciences | 30 | 55 | 3,207 | 14,209 |
| Civil engineering | 9 | 53 | 1,583 | 6,908 |
| Industrial and information engineering | 42 | 73 | 5,239 | 40,246 |
| Total | 192 | 86 | 36,450 | 206,433† |

\* The figure refers to publications (2008-2012) authored by at least one professor pertaining to the UDA.
† The total is less than the sum of the column data due to double counts of individual publications co-authored by professors that belong to different UDAs.

As for the territorial distribution of universities, we refer to the Nomenclature of Territorial Units for Statistics (NUTS)[5]. In Italy the aggregations provided under legislation for the national units of political and administrative decentralization are the Regions (NUTS 2) and Provinces (NUTS 3). The Italian state is subdivided in 20 regions and 110 provinces. Table 2 shows the regional distribution of Italian academia. There is at least one university in each region. In Valle d'Aosta, the local private university employs one professor only in the UDAs under investigation. In the following we then refer to 19 regions out of 20. Provinces with at least one university are 54 in all[6].

---

[2] http://cercauniversita.cineca.it/php5/docenti/cerca.php, last accessed on July 15, 2015.
[3] The complete list is accessible on http://attiministeriali.miur.it/UserFiles/115.htm, last accessed July 15, 2015.
[4] Mathematics and computer sciences; Physics; Chemistry; Earth sciences; Biology; Medicine; Agricultural and veterinary sciences; Civil engineering; Industrial and information engineering.
[5] NUTS is a geocode standard for referencing the subdivisions of countries for statistical purposes. The standard is developed and regulated by the European Union, and thus only covers the member states of the EU in detail.

*Table 2: The regional distribution of Italian academia*

| Region | Macro-area | Inhabitants (x 1,000) | Universities[§] | Total professors[*] | UDAs[†] |
|---|---|---|---|---|---|
| Abruzzo | South & islands | 1,323 | 3 | 915 | 9 |
| Basilicata | South & islands | 591 | 1 | 255 | 8 |
| Calabria | South & islands | 2,006 | 3 | 791 | 9 |
| Campania | South & islands | 5,806 | 5 | 3,394 | 9 |
| Emilia Romagna | Northeast | 4,284 | 4 | 3,618 | 9 |
| Friuli Venezia Giulia | Northeast | 1,222 | 3 | 967 | 9 |
| Lazio | Center | 5,534 | 8 | 4,490 | 9 |
| Liguria | Northwest | 1,612 | 1 | 982 | 8 |
| Lombardy | Northwest | 9,646 | 10 | 5,452 | 9 |
| Marche | Center | 1,549 | 4 | 843 | 9 |
| Molise | South & islands | 321 | 1 | 133 | 3 |
| Piedmont | Northwest | 4,395 | 3 | 2,094 | 9 |
| Puglia | South & islands | 4,076 | 4 | 1,825 | 9 |
| Sardinia | South & islands | 1,665 | 2 | 1,147 | 9 |
| Sicily | South & islands | 5,029 | 4 | 3,176 | 9 |
| Tuscany | Center | 3,675 | 5 | 3,158 | 9 |
| Trentino Alto Adige | Northeast | 1,007 | 2 | 276 | 6 |
| Umbria | Center | 884 | 1 | 839 | 9 |
| Valle d'Aosta | Northwest | 126 | 0 | 1 | 0 |
| Veneto | Northeast | 4,828 | 4 | 2,094 | 9 |
| Total | | 59,579 | 68 | 36,450 | 9 |

[*] *Number of professors belonging to one of the 192 SDSs under observation.*

[†] *Number of UDAs under observation with at least 10 professors.*

[§] *Number of universities in the region with at least 10 professors in the 192 SDS under observation.*

## 2.2 Measuring research productivity

Productivity is the quintessential indicator of efficiency in any production system. Our aim here is to measure first the research productivity of each professor and then average the productivities of all professors of the same field and region. When measuring labor productivity, if there are differences in the production factors (scientific instruments, materials, databases, support staff, etc.) available to each scientist then there should be normalization for them. Unfortunately, relevant data at the individual level are not available in Italy. Thus an often-necessary assumption is that the resources available to single researchers within the same field are the same. A further assumption, again unless specific data are available, is that the hours devoted to research are more or less the same for each individual.

However, because of the funding allocation system described above, we can assume that possible minor differences in the production factors should not cause notable distortions in productivity measures. Furthermore, universities are of different sizes and more or less disciplinary-focused, which may affect research productivity of individuals. However, it has been shown that in general there are no notably varying returns to scale (Abramo et al., 2012a) and to scope of research (Abramo et al., 2014a). To assess research productivity of individual professors we do not limit ourselves to a simple count of their indexed publications, but we consider their outcome, meaning the impact of each authored publication, over the five-year period observed. As a proxy of impact we adopt the number of citations observed at 31/05/2014. The citation window is large enough to assure an adequate estimate of the impact of each publication (Abramo et al., 2011). It is very possible that professors belonging to a particular scientific field will also publish outside that field. Because citation behavior varies across fields, we standardize the citations for each publication with respect to the average of the distribution of citations for all the cited Italian publications indexed in the same year and the same WoS subject category[7]. Furthermore, research projects frequently involve a team of scientists, which shows in co-authorship of publications, therefore we account for the fractional contribution of the scientists to output. Thus in formula, the proxy for yearly productivity of a single researcher, which we name Fractional Scientific Strength, *FSS*, is:

$$FSS = \frac{1}{t} \cdot \sum_{i=1}^{N} \frac{c_i}{\bar{c}_i} f_i$$

---

[6] Very few universities have schools localized in different provinces. In these cases we have localized the university in the province of the central administration.

Where:
$t$ = number of years of work by researcher in period under observation
$N$ = number of publications by researcher in period under observation
$c_i$ = citations received by publication $i$;
$\bar{c}_i$ = average of distribution of citations received for all cited publications in same year and subject category of publication $i$
$f_i$ = fractional contribution of the researcher to publication $i$.

Fractional contribution equals the inverse of the number of authors, in those fields where the practice is to place the authors in simple alphabetical order, but assumes different weights in other cases. For the life sciences, widespread practice in Italy and abroad is for the authors to indicate the various contributions to the published research by the order of the names in the byline. For these areas, we give different weights to each co-author according to their order in the byline and the character of the co-authorship (intra-mural or extra-mural). If first and last authors belong to the same university, 40% of citations are attributed to each of them; the remaining 20% are divided among all other authors. If the first two and last two authors belong to different universities, 30% of citations are attributed to first and last authors; 15% of citations are attributed to second and last but one author; the remaining 10% are divided among all others[8]. Failure to account for the number and position of authors in the byline would result in notable distortions as shown in Abramo et al. (2013a).

Because the intensity of publications varies across field (Garfield, 1979; Butler, 2007; Abramo and D'Angelo, 2007), in order to avoid distortions in productivity rankings at the aggregate levels (UDA) we then compare professors within the same field (Abramo et al., 2013). We calculate the productivity of each scientist in each SDS and express it as the ratio to the average productivity of all Italian productive professors of the same SDS, $FSS^N$ (normalized FSS). For more details on the assumptions and limits of the adopted research productivity measure we refer the reader to Abramo and D'Angelo (2014). Differently from other indicators of research performance, FSS embeds both quantity and impact of production, similarly to the h-index. However, differently from the h-index and most of its variants, it does not neglect the impact of works with a number of citations below h and all citations above h of the h-core works. It does not fail either to field-normalize citations, and to account for the number of co-authors and their order in the byline where appropriate.

---

[7] As shown by Abramo et al. (2012b), this scaling factor seems the most reliable for Italian data.
[8] The weighting values were assigned following advice from senior Italian professors in the life sciences and could be changed to suit different practices in other national contexts.

## 3. Ranking of territories by research productivity for each field

In this section we rank the territories by average research productivity of the professors in local universities, in each field and discipline. For significance reasons we exclude the territories with less than six professors in the given SDS, or 10 in the UDA. We begin the analyses at the higher territorial level of the regions and then proceed to the provinces.

### 3.1 Analysis at the regional level

The regional research productivity in a field (SDS) is calculated as the average FSS of all professors in the SDS employed at local universities. As an example, Table 3 presents the data for BIO/10 (Biochemistry). This is an SDS with almost 1,000 active professors, distributed in all Italian regions (19 in total). However, the regions of Molise and Trentino Alto Adige are excluded from the rankings, having less than six professors active in the SDS. For improved readability of the comparative performance, we show the normalized productivity $FSS^N$.

*Table 3: Ranking of regions by average research productivity of Biochemistry professors*

| Region | Research staff | $FSS^N$ | rank |
|---|---|---|---|
| Veneto | 64 | 1.295 | 1 |
| Tuscany | 67 | 1.176 | 2 |
| Piedmont | 35 | 1.117 | 3 |
| Lazio | 105 | 1.101 | 4 |
| Puglia | 47 | 1.084 | 5 |
| Lombardy | 146 | 1.060 | 6 |
| Emilia Romagna | 93 | 1.042 | 7 |
| Marche | 44 | 0.994 | 8 |
| Friuli Venezia Giulia | 22 | 0.952 | 9 |
| Abruzzo | 27 | 0.872 | 10 |
| Liguria | 20 | 0.864 | 11 |
| Campania | 103 | 0.685 | 14 |
| Sardinia | 25 | 0.656 | 15 |
| Sicily | 60 | 0.576 | 16 |
| Umbria | 26 | 0.507 | 17 |
| Calabria | 16 | 0.460 | 18 |
| Basilicata | 6 | 0.313 | 19 |

The territory with the highest average productivity of professors is Veneto: the region's 64 professors present an FSS that is 29.5% higher than the average achievement of their national colleagues. Immediately below Veneto we find Tuscany (+17.6%), Piedmont (+11.7%) and Lazio (+10.1%). At the bottom of the ranking are Basilicata, Calabria and Umbria. Contrary to most other SDSs, a significant correlation between size and productivity exists. There is also a strong concentration of the southern regions in the lower part of the national ranking. Which factor has a prevailing effect on the final ranking could be further investigated but it outside the scope of the current work. This analysis can be repeated for all the fields of interest (SDSs), showing where research investment offers the highest and lowest returns.

The same analysis can also be conducted at the discipline (UDA) level. As an example, in Table 4 we present the regional ranking of the professors operating in the 12 SDSs of Chemistry. As indicated in Section 2.2, we aggregate the individual productivities at the UDA level after normalizing them to the average of all productive professors of the same field (SDS). Molise and Trentino Alto Adige are absent from the ranking, having less than 10 professors active in the UDA. Tuscany, a large region for staff numbers, tops the ranking for productivity, followed by three 'middle-sized' regions (Calabria, Umbria and Piedmont). At the bottom of the ranking are Basilicata, the Marche and Liguria. Unlike the case of the BIO/10 field, the analysis of the Chemistry UDA does not show any association between productivity and geographic macro-area, since both at the top and the bottom of the ranking include a mix of regions from north, south and central Italy.

*Table 4: Ranking of regions by average research productivity of Chemistry professors*

| Region | Research staff | $FSS^N$ | rank |
|---|---|---|---|
| Tuscany | 308 | 1.424 | 1 |
| Calabria | 74 | 1.228 | 2 |
| Umbria | 92 | 1.094 | 3 |
| Piedmont | 189 | 1.047 | 4 |
| Emilia Romagna | 472 | 1.033 | 5 |
| Lazio | 252 | 0.939 | 6 |
| Campania | 278 | 0.929 | 7 |
| Friuli Venezia Giulia | 72 | 0.929 | 7 |
| Lombardy | 363 | 0.925 | 9 |
| Veneto | 206 | 0.898 | 10 |
| Sicily | 303 | 0.827 | 11 |
| Puglia | 147 | 0.780 | 12 |
| Abruzzo | 52 | 0.772 | 13 |
| Sardinia | 135 | 0.767 | 14 |
| Liguria | 89 | 0.737 | 15 |
| Marche | 96 | 0.625 | 16 |

**3.2 Analysis at the provincial level**

The analysis at the regional level can inform the research and industrial policies of national governments, as well as of the regions. The provincial level analysis should again be useful to regional administrators, but also to private actors seeking the geographic locations of the more productive research groups in particular fields of interest. Figure 1 provides examples of the analysis at provincial level, for both the individual fields and the broader scientific disciplines:
- in the left graph, the distribution of average $FSS^N$ for professors belonging to SDS BIO/10 (Biochemistry);

- in the right graph, the distribution of average $FSS^N$ for professors belonging to the Chemistry UDA.

The two analyses refer to different levels of aggregation and provide different types of information. For example there are territories indicated as "no data", where scientific activity in the SDS or UDA is completely absent, or limited to very small research groups. The intensity of the shading provides detailed information on the levels of productivity. For example in the case of the Chemistry UDA (right graph) we can observe a north-south "dorsal", with highly productive provinces running down the backbone of the country. In the case of the Biochemistry SDS (left graph), the areas of excellence in the south are limited to the Province of Lecce (the Italian 'heel'), while there are substantial clusters of excellence in central-northern and central Italy (respectively the provinces of Florence, Bologna and Modena; the provinces of Rome and L'Aquila).

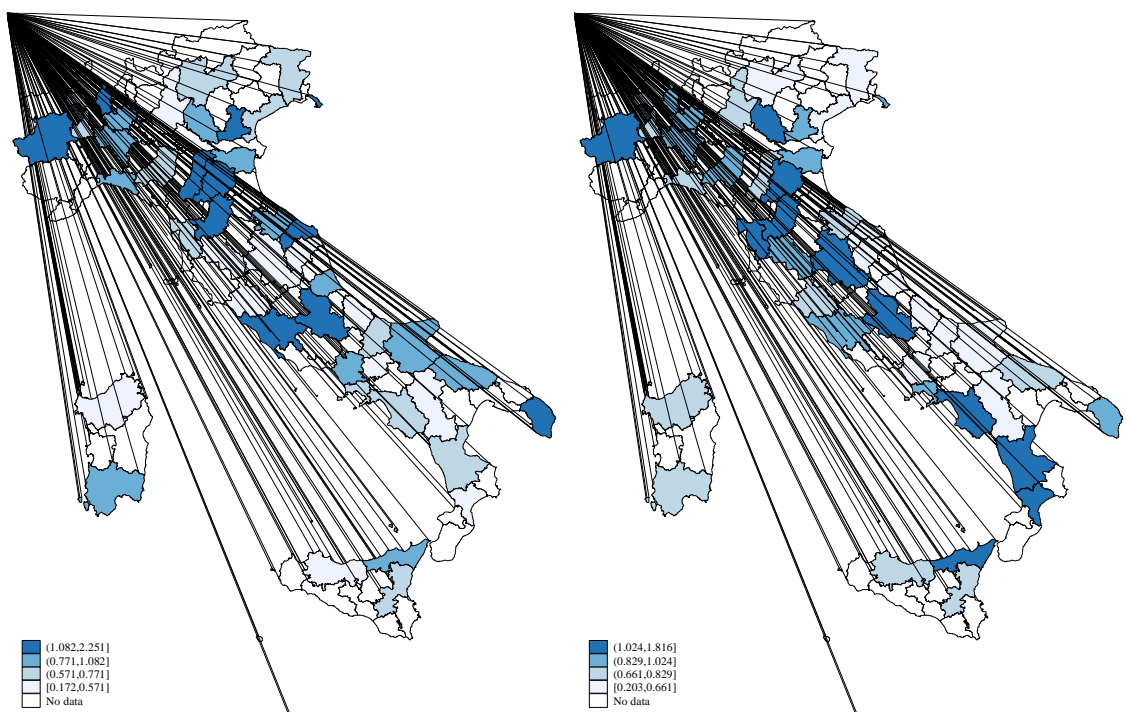

*Figure 1: Distribution of average productivity per province of professors in the Biochemistry field (left); of professors in the Chemistry discipline (right)*

## 4. Ranking the scientific fields of a given territory by research productivity

To complement the analysis just presented we can evaluate the performance of the research fields within the individual region or province, to answer the very pertinent question: for a given territory, which are the fields or disciplines where the research productivity is highest or lowest? This analysis concerns the regional governments, interested in defining policies to build on local strengths and reinforce less-productive fields strategic to territorial development. Once again we begin the analysis at the regional level and proceed to the finer provincial level.

**4.1 Analysis at the regional level**

As an example, in Table 5 we present the analysis for Campania region: second in Italy for population and fourth for university faculty, including 3,394 professors in the SDSs investigated. These belong to 181 SDSs, of which 156 have at least six professors. For limited space, Table 5 presents only the first and last 10 SDSs of the productivity ranking. At the top is MED/03 (Medical genetics), showing an $FSS^N$ at 2.951, an average productivity almost triple the national average for that field. Immediately following we find BIO/15 (Pharmaceutic biology), with average productivity at 2.355, and VET/08 (Clinical veterinary medicine) at 2.134. Besides VET/08, the UDA of Agricultural and veterinary sciences (UDA 7) places another two SDSs among the top 10 of the region: VET/03 (General pathology and veterinary pathological anatomy) and AGR/13 (Agricultural chemistry). Also in the top-ten for productivity are two Chemistry SDSs and one form (Chemical)-Industrial Engineering (ING-IND/26), as well as two SDSs of Civil engineering. At the bottom of the ranking we find four SDSs of Industrial engineering and the same number from Medicine, as well as ICAR/06 (Topography and cartography) and MAT/04 (Complementary mathematics).

*Table 5: Top 10 and bottom 10 SDSs by research productivity as compared to other SDSs in Campania region*

| SDS | UDA | $FSS^N$ | National Percentile* |
|---|---|---|---|
| MED/03-Medical genetics | 6 | 2.951 | 100 |
| BIO/15-Pharmaceutic biology | 5 | 2.355 | 93.3 |
| VET/08-Clinical veterinary medicine | 7 | 2.134 | 100 |
| CHIM/10-Food chemistry | 3 | 2.012 | 100 |
| ING-IND/26-Theory of development for chemical processes | 9 | 1.985 | 90.0 |
| AGR/13-Agricultural chemistry | 7 | 1.806 | 93.3 |
| CHIM/04-Industrial chemistry | 3 | 1.800 | 100 |
| ICAR/04-Road, railway and airport construction | 8 | 1.707 | 100 |
| VET/03-General pathology and veterinary pathological anatomy | 7 | 1.601 | 91.7 |
| ICAR/03-Environmental and health engineering | 8 | 1.571 | 92.3 |
| … | - | - | - |
| ING-IND/35-Engineering and management | 9 | 0.150 | 7.7 |
| ICAR/06-Topography and cartography | 8 | 0.148 | 37.5 |
| MED/41-Anaesthesiology | 6 | 0.138 | 6.7 |
| ING-IND/15-Design and methods for industrial engineering | 9 | 0.105 | 0 |
| MED/33-Locomotory diseases | 6 | 0.094 | 20.0 |
| ING-IND/17-Industrial and mechanical plant | 9 | 0.089 | 0 |
| MED/31-Otorinolaringology | 6 | 0.084 | 0 |
| MED/43-Legal medicine | 6 | 0.078 | 6.3 |
| MAT/04-Complementary mathematics | 1 | 0.059 | 50.0 |
| ING-IND/02-Naval and marine construction and installation | 9 | 0.040 | 0 |

*\* 100 = top*

However, referring again to the least productive fields seen in Table 5, we observe an interesting phenomenon: although the professors of MAT/04 are last but one by productivity among the SDSs of their region, they show exactly median performance for this SDS at the national level (i.e. compared to all other regions). It could thus be interesting to answer another question: for each territory, what are the fields of research where the productivity of professors is high or low in the national perspective?

Table 6 provides the answer to this question, in the listing of the best and worst 10 SDSs for the case of Campania, however this time by national percentile of FSS, meaning in comparison to other regions. The list does not superimpose on that of Table 5, although there is a strong correlation (Spearman ρ-value of 0.91 for the 156 observations). Seven out of the top 10 SDSs appear in both rankings (Table 5 and 6), however there are substantial changes among the bottom 10 SDSs, where there are five new entries. Medicine and Industrial engineering continue with strong representation among the least-performing (four SDSs each). Table 5 and Table 6 answer two different questions. Table 5 presents the SDSs in the region with the higher (lower) returns on research investments and is aimed at informing selective funds allocation. Table 6 shows the SDSs where the region has a research competitive advantage vis-a-vis the other regions and is aimed at informing strategic decisions.

*Table 6: Top 10 and bottom 10 SDSs by research productivity in the Region of Campania, in terms of their national ranking*

| SDS | UDA | $FSS^N$ | National percentile* |
|---|---|---|---|
| CHIM/04-Industrial Chemistry | 3 | 1.800 | 100 |
| CHIM/10-Food Chemistry | 3 | 2.012 | 100 |
| ICAR/04-Road, Railway and Airport Construction | 8 | 1.707 | 100 |
| ING-IND/25-Chemical Plants | 9 | 1.428 | 100 |
| MED/03-Medical Genetics | 6 | 2.951 | 100 |
| VET/08-Clinical Veterinary Medicine | 7 | 2.134 | 100 |
| ING-IND/11-Environmental Technical Physics | 9 | 1.249 | 94.4 |
| BIO/03-Environmental and Applied Botanics | 5 | 1.454 | 94.1 |
| AGR/13-Agricultural Chemistry | 7 | 1.806 | 93.3 |
| BIO/15-Pharmaceutic Biology | 5 | 2.355 | 93.3 |
| … | - | - | - |
| MED/43-Legal Medicine | 6 | 0.078 | 6.3 |
| BIO/05-Zoology | 5 | 0.239 | 0 |
| CHIM/01-Analytical Chemistry | 3 | 0.276 | 0 |
| ING-IND/01-Naval Architecture | 9 | 0.181 | 0 |
| ING-IND/02-Naval and Marine construction and installation | 9 | 0.040 | 0 |
| ING-IND/15-Design and Methods for Industrial Engineering | 9 | 0.105 | 0 |
| ING-IND/17-Industrial and Mechanical Plant | 9 | 0.089 | 0 |
| MED/15-Blood Diseases | 6 | 0.210 | 0 |
| MED/16-Rheumatology | 6 | 0.246 | 0 |
| MED/31-Otorinolaringology | 6 | 0.084 | 0 |

*\* 100 = top*

An interesting aspect to be grasped from both tables is the remarkable heterogeneity of performance among the SDSs of a given UDA. For the example of Campania, SDSs from both Medicine and Industrial engineering appear both at the top and the bottom of both lists. Thus it could be useful to evaluate the productivity of the regional research fields, but aggregated at the level of the disciplinary area, to understand how much this heterogeneity in fact influences the positioning of the regional UDAs at the national level. Table 7 presents the results of this type of analysis: each cell presents the average value of $FSS^N$ for the professors of a given region and UDA. The table permits response to both the above questions, since it can be analyzed from two perspectives: reading the data by row, we can compare the productivity of a UDA to that of all the others active in the same region. Reviewing the data by column, we can instead arrive at the comparative evaluation of a region's productivity in a given UDA, compared to all the other regions active in the same UDA.

*Table 7: Research productivity (FSS) of academics by UDA and region*

| Region | UDA* | | | | | | | | |
|---|---|---|---|---|---|---|---|---|---|
| | 1 | 2 | 3 | 4 | 5 | 6 | 7 | 8 | 9 |
| Abruzzo | 0.493 | 0.691 | 0.772 | 0.672 | 0.645 | 0.647 | 0.828 | 0.283 | 0.560 |
| Basilicata | 0.828 | 0.819 | 0.447 | 0.687 | 0.431 | - | 0.527 | 0.290 | 0.483 |
| Calabria | 0.989 | 1.380 | 1.228 | 0.878 | 0.885 | 0.777 | 0.365 | 0.676 | 0.942 |
| Campania | 0.639 | 0.802 | 0.929 | 0.850 | 0.720 | 0.637 | 0.948 | 0.948 | 0.833 |
| Emilia Romagna | 0.736 | 0.695 | 1.033 | 0.832 | 1.059 | 0.998 | 0.854 | 1.022 | 0.940 |
| Friuli Venezia Giulia | 1.058 | 1.062 | 0.929 | 0.768 | 1.097 | 0.858 | 0.587 | 0.502 | 0.703 |
| Lazio | 0.720 | 0.929 | 0.939 | 1.130 | 0.970 | 0.623 | 1.124 | 0.591 | 0.734 |
| Liguria | 0.507 | 0.810 | 0.737 | 0.701 | 0.898 | 0.943 | - | 0.681 | 0.621 |
| Lombardy | 0.863 | 1.159 | 0.925 | 1.139 | 1.062 | 1.194 | 0.959 | 0.773 | 0.880 |
| Marche | 0.314 | 0.830 | 0.625 | 0.741 | 0.935 | 0.870 | 0.643 | 0.830 | 0.708 |
| Molise | - | - | - | - | 0.585 | 0.526 | 0.542 | - | - |
| Piedmont | 0.740 | 0.884 | 1.047 | 0.695 | 1.419 | 1.189 | 1.069 | 1.038 | 0.923 |
| Puglia | 0.627 | 1.073 | 0.780 | 0.648 | 0.836 | 0.639 | 1.009 | 0.593 | 0.788 |
| Sardinia | 0.453 | 0.585 | 0.767 | 0.404 | 0.590 | 0.424 | 0.470 | 0.325 | 0.693 |
| Sicily | 0.978 | 0.697 | 0.827 | 0.646 | 0.571 | 0.458 | 0.608 | 0.569 | 0.711 |
| Tuscany | 0.764 | 0.897 | 1.424 | 1.054 | 0.991 | 1.023 | 0.642 | 0.391 | 0.797 |
| Trentino Alto Adige | 1.117 | 1.423 | - | - | 0.827 | - | 0.972 | 1.235 | 1.542 |
| Umbria | 1.048 | 0.930 | 1.094 | 0.645 | 0.596 | 0.781 | 0.577 | 0.943 | 0.966 |
| Veneto | 0.742 | 0.991 | 0.898 | 1.392 | 1.190 | 1.360 | 1.208 | 0.865 | 1.174 |

*\* 1 = Mathematics and computer sciences; 2 = Physics; 3 = Chemistry; 4 = Earth sciences; 5 = Biology; 6 = Medicine; 7 = Agricultural and veterinary sciences; 8 = Civil engineering; 9 = Industrial and information engineering*

**4.2 Analysis at the provincial level**

For the analysis of the productivity of research fields active in a province, we take the example of Rome: a province of 13 universities, with 4,206 professors in the SDSs under examination, a full 96.4% of which are concentrated in three generalist institutions (the universities 'La Sapienza', 'Tor Vergata' and 'Roma Tre')[9]. The 4206 academics are structured in 162 SDSs, of which 138 have over five professors. Table 8 presents the first and last 10 SDSs for average productivity.

In the top three of the list we see two SDSs of UDA 4 (Earth sciences): first is GEO/03, with an average $FSS^N$ of 2.7, and third is GEO/06 (Mineralogy), with average $FSS^N$ of 1.9. Near the very top of the rankings are also two Biology SDSs (BIO/03 and BIO/06). BIO/03 faculty also has an average productivity more than double that of any other province in Italy. The top 10 also includes 4 SDSs in Medicine (MED/10, MED/33, MED/19 and MED/21), with average $FSS^N$ never below 1.4. The Aerospatial propulsion SDS (ING-IND/07) deserves a special note: as one of the most productive faculties of the province (and also at the national level, as indicated by the last column), it reflects the importance of a sector that is also strategic from the industrial point of view, for this particular territory.[10]

---

[9] The remaining 3.6% of the academic staff in the Province of Rome are employed in private universities, primarily active in the social sciences, arts and humanities, many of which operate by Internet..

[10] The Province of Rome hosts a significant cluster of large, medium and small enterprises active in the aerospace sector.

The bottom 10 of the ranking list consists entirely of SDSs with average $FSS^N$ always below 0.4: five of these are SDSs of Industrial engineering, and two each are from Civil engineering and Medicine. The very last of the list is MAT/04 (Complementary mathematics). However, the value shown in the last column of Table 8 indicates that in reality the Rome faculty for this SDS show an average performance that places them in the first national quartile, when compared to the SDSs of all the provinces. For this, as at the regional level of the analysis, in Table 9 we present the list of the top and bottom 10 SDSs by research productivity in Rome province, as compared to other provinces: an approach useful for understanding what are the Rome fields of research where the productivity of professors is higher or lower, from a national perspective. Comparing the two lists (Table 8 to 9), there are only three new entries in the top ten (BIO/08, ING-IND/05 and ING-IND/35), however in the lower part of the list there are a full eight SDSs that no longer appear among the 'worst-performing' (only ING-IND/10 and ING-IND/14 are present in both the lists).

*Table 8: Top 10 and bottom 10 SDSs in the Province of Rome by research productivity, compared to other SDSs in the same province*

| SDS | UDA | FSS$^N$ | National percentile |
|---|---|---|---|
| GEO/03-Structural Geology | 4 | 2.696 | 100 |
| BIO/03-Environmental and Applied Botanics | 5 | 2.234 | 100 |
| GEO/06-Mineralogy | 4 | 1.885 | 94.7 |
| MED/10-Respiratory Diseases | 6 | 1.594 | 86.4 |
| MED/33-Locomotory Diseases | 6 | 1.585 | 92.9 |
| CHIM/02-Physical Chemistry | 3 | 1.494 | 93.8 |
| BIO/06-Comparative Anatomy and Citology | 5 | 1.483 | 83.9 |
| MED/19-Plastic Surgery | 6 | 1.472 | 90.0 |
| ING-IND/07-Aerospatial Propulsion | 9 | 1.435 | 100.0 |
| MED/21-Thoracic Surgery | 6 | 1.417 | 78.6 |
| … | - | - | - |
| MED/01-Medical Statistics | 6 | 0.382 | 31.6 |
| ICAR/09-Construction Techniques | 8 | 0.372 | 29.4 |
| MED/14-Nephrology | 6 | 0.362 | 38.1 |
| ING-IND/17-Industrial and Mechanical Plant | 9 | 0.361 | 46.2 |
| ICAR/07-Geotechnics | 8 | 0.358 | 46.2 |
| ING-IND/32-Electrical Convertors, Machines and Switches | 9 | 0.344 | 26.7 |
| ING-IND/28-Excavation Engineering and Safety | 9 | 0.306 | 33.3 |
| ING-IND/14-Mechanics and Machine Design | 9 | 0.189 | 14.8 |
| ING-IND/10-Technical Physics | 9 | 0.148 | 7.4 |
| MAT/04-Complementary Mathematics | 1 | 0.141 | 76.2 |

*Table 9: Top 10 and bottom 10 SDSs in the Province of Rome by research productivity, as compared to other provinces*

| SDS | UDA | FSS$^N$ | National percentile |
|---|---|---|---|
| BIO/03-Environmental and Applied Botanics | 5 | 2.234 | 100 |
| BIO/08-Anthropology | 5 | 1.339 | 100 |
| GEO/03-Structural Geology | 4 | 2.696 | 100 |
| ING-IND/05-Aerospace Systems | 9 | 1.029 | 100 |
| ING-IND/07-Aerospatial Propulsion | 9 | 1.435 | 100 |
| ING-IND/35-Engineering and Management | 9 | 1.080 | 95.8 |
| GEO/06-Mineralogy | 4 | 1.885 | 94.7 |
| CHIM/02-Physical Chemistry | 3 | 1.494 | 93.8 |
| MED/33-Locomotory Diseases | 6 | 1.585 | 92.9 |
| MED/19-Plastic Surgery | 6 | 1.472 | 90.0 |
| … | - | - | - |
| CHIM/09-Applied Technological Pharmaceutics | 3 | 0.666 | 25.9 |
| CHIM/03-General and Inorganic Chemistry | 3 | 0.442 | 23.7 |
| ING-INF/07-Electric and Electronic Measurement Systems | 9 | 0.483 | 23.1 |
| MAT/08-Numerical analysis | 1 | 0.475 | 22.6 |
| MED/06-Medical Oncology | 6 | 0.470 | 21.7 |
| CHIM/04-Industrial Chemistry | 3 | 0.508 | 18.8 |
| MED/09-Internal Medicine | 6 | 0.449 | 18.2 |
| ING-IND/14-Mechanics and Machine Design | 9 | 0.189 | 14.8 |
| GEO/08-Geochemistry and Volcanology | 4 | 0.514 | 11.8 |
| ING-IND/10-Technical Physics | 9 | 0.148 | 7.4 |

## 5. Conclusions

The theme of territorial mapping of research activity is active in the literature, because of its potential utility to different stakeholders. First of all, mapping of research assists national policy-makers, in the development of coherently organized overall R&D policies. It also assists regional policy-makers, responsible for promoting local development. Finally, it assists private companies, for example in making choices in the localization of R&D activities, or among potential collaborations with public research institutions.

The scientific profiling of territories has until now been conducted by bibliometric means that analyze the geographic distribution of scientific output through two types of indicators: i) those that examine 'production' (such as number of publications or citations, share of national/world publications/citations, or share of top cited publications), but which are typically size dependent; ii) those that attempt to normalize the values of such indicators respect to some factor of size. This factor can be: a) the production itself (average impact of publications, concentration index of top cited publications, indices of sectorial specialization, etc.), or b) some macroeconomic feature of the territory (population, GDP, etc.).

To the authors' knowledge, the current proposal represents the first attempt to measure the research efficiency (i.e. the productivity of territories) by means of the impact produced by every individual professor located in the territory. Notwithstanding the limits and assumptions embedded in its calculation, and the inevitable uncertainty level associated to its value, research productivity measures are very important, since for the potential research funders, they reveal where research spending has the highest (or lowest) returns. The proposed measurement begins from the individual academics working in Italian universities as the unit of observation. It is typical that academic research is the principle component of public research infrastructure: at the Italian level, the academic sphere represents 2/3 of the total research staff. The measurement provides precise information concerning the scientific production of each professor, indexed in the WoS for the period 2008-2012, as well as their specific field of scientific investigation.

The proposed analysis simultaneously reveals: i) for each territory, the fields of research where the productivity is highest or lowest and ii) for each scientific field, which are the territories with the highest or lowest productivity. For both outlooks, the analysis can examine two levels: the 'macro' level of the region, and the more detailed level of the province. With these different approaches, the various stakeholders can then obtain the information required to support their different types of decisions.

The replicability of the analysis in other countries is conditioned by the availability of detailed data on the distribution of the research staff in the territory, as well as the fields of research in which the individuals operate. Our hope is that other scholars can conduct similar analyses, providing information for international comparison, even if this is limited to specific fields.

The mapping realized, presented in this paper in a limited manner, can also prompt analyses by our national colleagues, to delve into the causes of productivity differences, which can arise from historical, sociological or structural macro-economic reasons.